\documentclass[11pt,preprint]{aastex}
\pdfoutput=1
\newcommand{\simgt}{\hbox{\rlap{\raise 0.425ex\hbox{$>$}}\lower 0.65ex\hbox{$\sim$}}}
\newcommand{\simlt}{\hbox{\rlap{\raise 0.425ex\hbox{$<$}}\lower 0.65ex\hbox{$\sim$}}}

\shorttitle{Statistical Mechanics of Collisionless Orbits. II.}
\shortauthors{Williams \& Hjorth}

\begin{document}

\title{
Statistical mechanics of collisionless orbits. II. Structure of halos}


\author{ 
Liliya L. R. Williams\altaffilmark{1,2,3}\email{llrw@astro.umn.edu}
Jens Hjorth\altaffilmark{2}\email{jens@dark-cosmology.dk}
}

\altaffiltext{1}{Department of Astronomy, University of Minnesota, 
116 Church Street SE, Minneapolis, MN 55455, USA}
\altaffiltext{2}{Dark Cosmology Centre, Niels Bohr Institute, 
University of Copenhagen, Juliane Maries Vej 30, DK-2100 Copenhagen \O, 
Denmark} 
\altaffiltext{3}{Institute for Theoretical Physics, University of Z\"urich, 
Winterthurerstrasse 190, CH-8057 Z\"urich, Switzerland}

\textheight8.95in
\voffset-0.45in
\pagestyle{plain} 


\begin{abstract}
In this paper, we present the density, $\rho$, velocity dispersion, $\sigma$, and 
$\rho/\sigma^3$ profiles of isotropic systems which have the energy distribution, 
$N(\varepsilon)\propto[\exp(\phi_0-\varepsilon)-1]$, derived in Paper I. This 
distribution, dubbed DARKexp, is the most probable final state of a collisionless 
self-gravitating system, which is relaxed in terms of particle energies, but not 
necessarily in terms of angular momentum. We compare the DARKexp predictions with 
the results obtained using the extended secondary infall model (ESIM). The ESIM numerical 
scheme is optimally suited for the purpose because (1) it relaxes only through 
energy redistribution, leaving shell/particle angular momenta unaltered, and 
(2) being a shell code with radially increasing shell thickness it has very good 
mass resolution in the inner halo, where the various theoretical treatments give
different predictions. The ESIM halo properties, and especially their energy 
distributions, are very well fit by DARKexp, implying that the techniques of 
statistical mechanics can be used to explain the structure of relaxed 
self-gravitating systems.
\end{abstract}

\section{Introduction}

Non-relativistic gravitational problems have been the subject of numerous studies, going back
to Newton. Despite the simplicity of the force law, these problems vary greatly in the degree 
of difficulty. The two-body problem is easy and the three-body problem has solutions if certain 
simplifying assumptions are made, while higher $N$ problems have no useful analytical solutions.  
When $N$ is large, one can distinguish two types of problems. In the many-body problem, the 
potential is grainy on scales comparable to the particle separation, while in the infinitely 
many body problem the potential is smooth on particle scales but can exhibit larger scale 
fluctuations. It is this last problem which is relevant to the collisionless stellar and dark 
matter systems, and in this series of papers we concentrate on this problem. 

To solve it, one can either numerically integrate individual particle orbits and measure the 
properties of the evolved system or appeal to the large-$N$ aspect of the problem and 
attempt to deduce the global properties of the systems, for example, the distribution function. 
The first approach is taken by the N-body simulations, which in the last couple of decades have 
produced consistent and robust results for the structure of collisionless dark matter halos
(see \cite{nav04} and \cite{sm09}, and references therein). These results are widely used in the 
literature whenever the properties of dark matter halos are called for. Examples of the second, 
mostly analytical approach can be found in \cite{lb67,binney82,sb87,mtj89,hm91} and \cite{sh92}.
While these do reproduce the general features of collisionless systems, they often use the 
knowledge of the final system (for example, the de Vaucouleurs profile of ellipticals) to 
motivate the choice of the distribution function. Since simulations make no such assumptions, 
and have convincingly demonstrated that the end result of collisionless relaxation is the universal 
profile, is an analytical derivation even needed?  We argue that it is.

In this series of papers, we develop an analytical approach, starting from first principles
which predicts the properties of relaxed self-gravitating collisionless systems, and test
it against numerically evolved systems.  In Paper I \citep{hw10} we showed that if one uses 
the techniques of statistical mechanics, and works in energy, or orbit space, instead of the 
usual phase-space, then the standard entropy maximization procedure yields the most probable 
state described by an exponential differential energy distribution, 
$N(\varepsilon)\propto\exp(\phi_0-\varepsilon)$, where $\varepsilon$ and $\phi_0$ are the 
dimensionless (positive) energy and the central potential depth, respectively. 
For finite potential depths, $N$ has to be truncated so that $N(\varepsilon>\phi_0)=0$.
An abrupt truncation will lead to an unphysical central density hole; any gradual transition 
to zero will lead to small integer occupation numbers in energy cells close to $\phi_0$.
(\cite{m96} addressed a similar problem in the case of the distribution function $f(\varepsilon)$.)
Because the standard Stirling approximation breaks down at low occupation numbers, we replace 
it with a superior approximation. With this, the most probable energy distribution becomes 
$N(\varepsilon)\propto[\exp(\phi_0-\varepsilon)-1]$, which we call DARKexp. It has only one 
free parameter, $\phi_0$, the depth of the system's central potential, or the energy of the 
most bound particle.

In this paper we derive the structural and dynamical properties of these systems, and compare
them to the results of a numerically evolved system. 

\section{Density profiles from DARKexp}\label{DARKexpden}

\begin{figure}[t]
\plotone{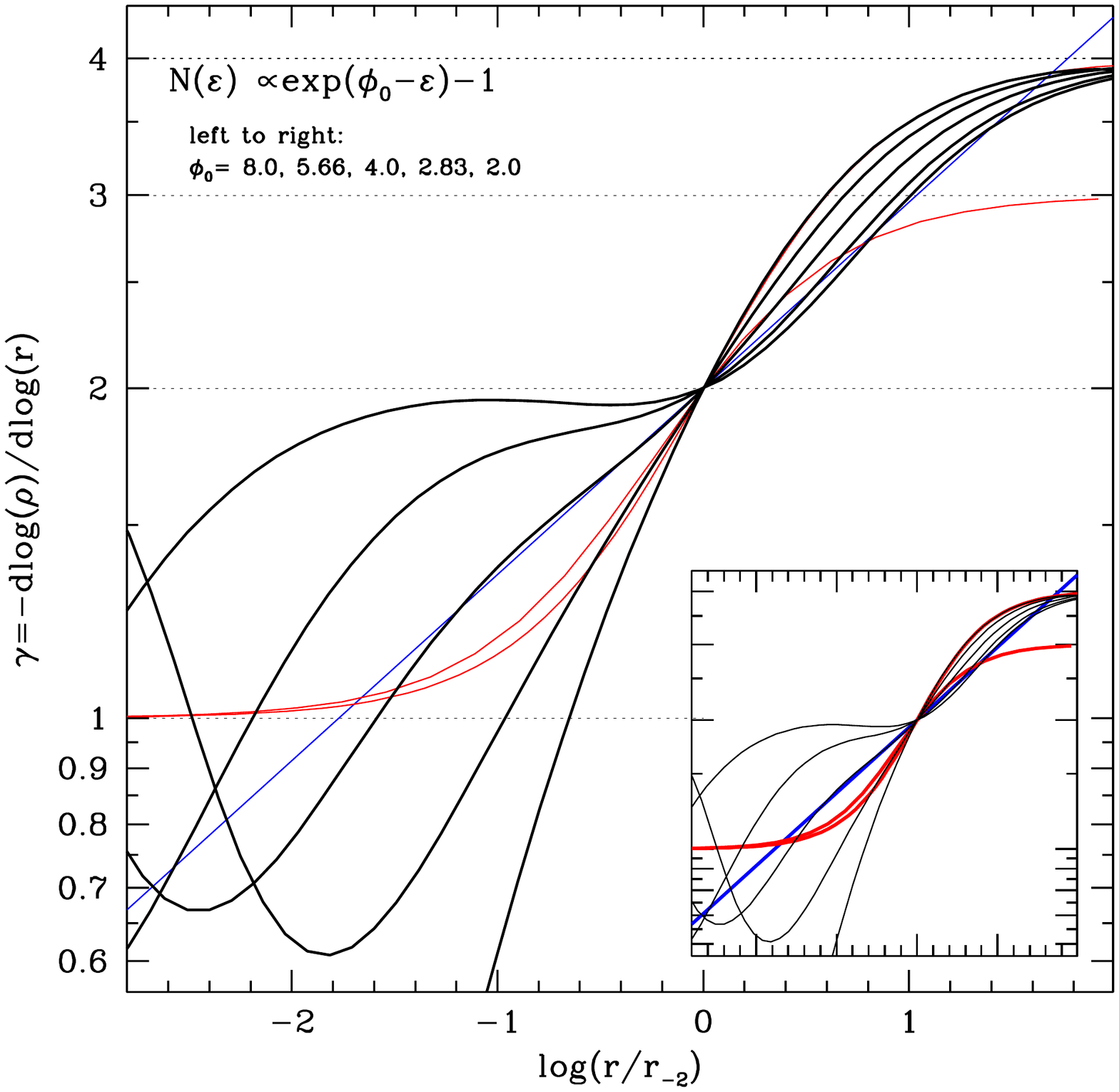}
\vskip-1.5in
\caption{Density profile slopes predicted by the isotropic DARKexp model (thick black lines).
From left to right the halo potential depths are $\phi_0=8.0, 5.66, 4.0, 2.83, 2.0$. These
five values span the range of typical behavior, and are spaced equally in log.
Note that the vertical scale is logarithmic, because the \cite{nav04} fits to 
N-body halos (Einasto profiles) are straight lines in this representation; $\alpha=0.17$ 
profile of \cite{nav04} is plotted as a straight blue line. The two red curves are the 
Hernquist and NFW profiles (NFW asymptotes to the outer slope of 3) shown for reference. 
The horizontal axis is normalized to the radius, $r_{-2}$, where the density slope attains 
$\gamma=2$. The inset is the same as the main figure, but plots DARKexp as thin lines so 
that the other three profiles are more visible.} 
\label{slopesALL}
\end{figure}

\begin{figure}[t]
\plotone{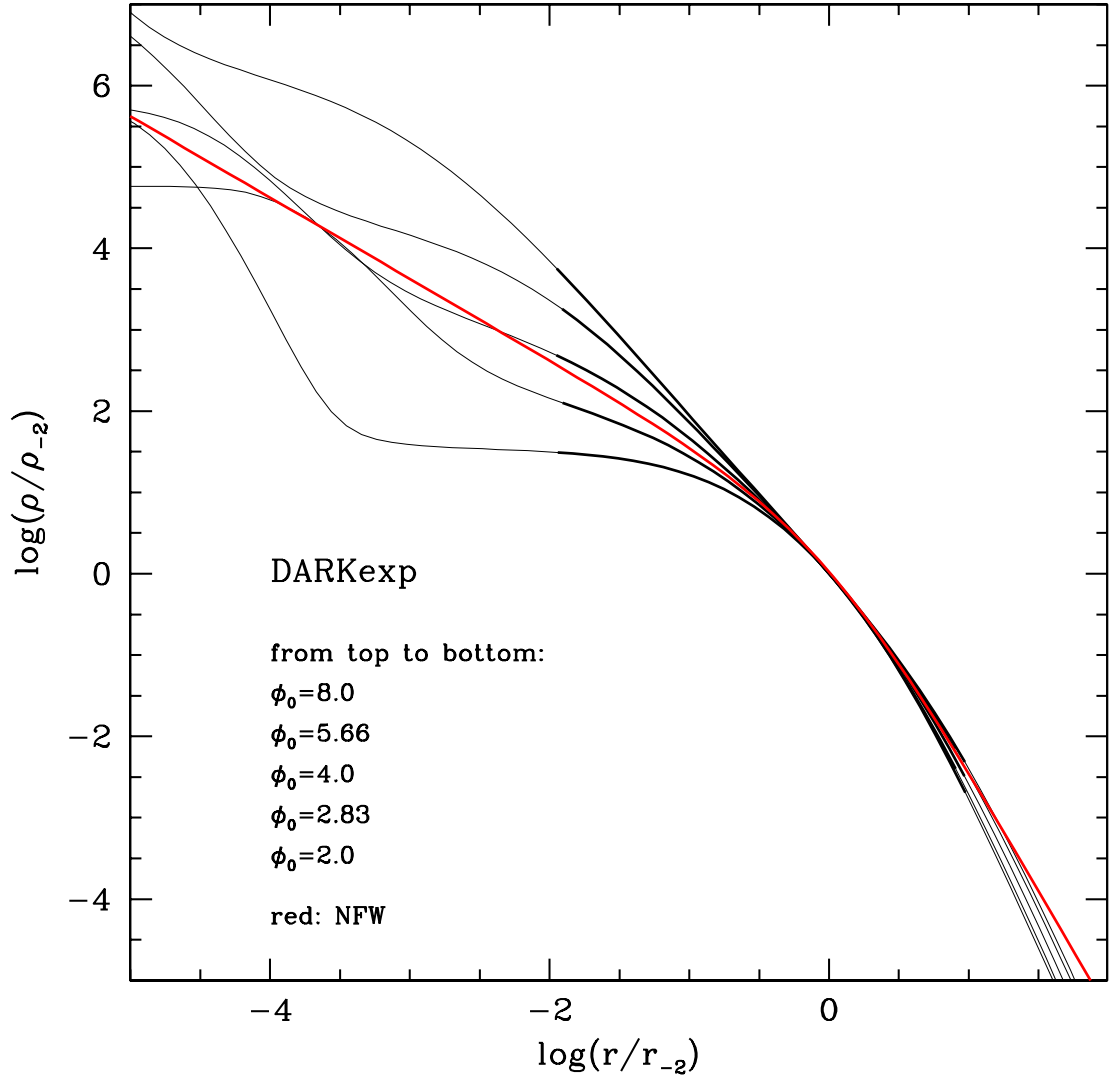}
\vskip-1.5in
\caption{Density profiles of the five DARKexp models shown in Figure~\ref{slopesALL}, normalized
such that at the radius where the density slope $\gamma=2$, the density is unity. The radial
range accessible to the highest resolution pure dark matter N-body simulations is highlighted
as thick lines. The red line is the NFW model shown for reference.}
\label{density2}
\end{figure}

Given a distribution function $f(E)$ one can integrate it over the velocity space to
obtain the density distribution as a function of radius, $\rho(r)$. However, given
a distribution in energy, $N(E)$ one cannot obtain $\rho(r)$ in one step. This is because
unlike $f(E)$, $N(E)$ contains in it the density of states, and so it already depends on
the potential, which is what one is trying to recover. Therefore an iterative procedure is 
needed, such as the one described in  \cite{binney82}, which assumes isotropy. 
Since there is only one free parameter in DARKexp, $\phi_0$, the density profiles are a 
function of $\phi_0$ only; these are shown in Figure~\ref{slopesALL}, where we plot the 
logarithmic density slope, $\log(\gamma)$, where $\gamma=-d\log(\rho)/d\log(r)$. 
The concept of `virial radius' is not defined in these systems, so we chose to normalize 
the radius to that where the density attains $\gamma=2$, and call it $r_{-2}$. The blue 
line is the \cite{nav04} $\alpha=0.17$ model, while the two red lines are the NFW 
\citep{nfw97}, and the Hernquist \citep{h90} profiles, shown for reference. The 
corresponding density profiles are plotted in Figure~\ref{density2}; thick lines 
highlight the radial range accessible to high-resolution numerical simulations 
\citep{nav04,sm09}. The red curve is the NFW profile, shown for comparison.

The DARKexp family of density profiles has a characteristic shape. At radii larger than
$\sim 0.01\; r_{-2}$, i.e. those that can be probed in simulated systems, and for small values 
of $\phi_0$ the density profiles get shallower monotonically from large to small radii, but 
for $\phi_0$ above $\sim 4$ the profiles first get shallower until they reach $\gamma\approx 2$, 
then stay around $2$ for a finite radial range before becoming shallow again. This type 
of non-monotonic slope behavior is known to occur in some analytical systems, for example,  
the isothermal sphere (\cite{BT87}, Section 4.4, Figure 4.8), polytropic systems with index 
$n$ around 5 \citep{mr01}, and systems where $\rho/\sigma^3$ is a power law in radius 
\citep{austin05}. In the case of DARKexp, the slope of 2 is `inherited' from the pure 
exponential $N(\varepsilon)=N_0 \exp(-\beta\varepsilon)$ studied by \cite{binney82}.

Paper I demonstrated analytically that DARKexp models must asymptote to a central density 
cusp of $\gamma=1$. Figures~\ref{slopesALL} and \ref{density2} show that inside of 
$\sim 0.1-0.01\; r_{-2}$ ~DARKexp density profiles undergo oscillations in slope. Though  
the amplitude of oscillations can be large, especially for systems with shallow potentials, 
they bracket the slope of 1. More significantly, because these oscillations involve very 
little mass and take place in a narrow linear radial interval, any physical system, such 
as an N-body halo, which exhibits slight deviations from spherical symmetry or has a small 
amount of substructure, would erase these oscillations, averaging them to the asymptotic slope. 
In other words, because the amount of mass involved is small, the difference between the
differential energy distribution of oscillating and a similar, but non-oscillating system 
will be small. (We note that some numerically generated systems do show small oscillations, 
see for example, Figure~2 of \cite{sm09} and Figure~7 of \cite{lud10}.) 

The DARKexp systems that happen to have the smallest oscillations, even before the smearing
effect of ellipticity or substructure is considered, are those with $\phi_0$ around 4, and 
so the asymptotic slope is already evident at radii somewhat interior to $r_{-2}$.  Curiously, 
DARKexp models that most resemble N-body profiles are also those with $\phi_0\approx 4$ 
(compare to the red NFW and Hernquist profiles, and the blue 
Einasto, or \cite{nav04} profiles, in Figure~\ref{slopesALL}).

\section{Testing DARKexp predictions}\label{testing}

\subsection{Choosing the appropriate numerical scheme}\label{chosing}

Because DARKexp makes a definite, one-parameter prediction about the shape of the energy
distribution of isotropic systems, and because the corresponding density profiles have
a distinct shape, the DARKexp model can be tested, using an appropriate numerical scheme. 
The scheme should satisfy certain criteria. 
\begin{enumerate}
\item It should evolve a self-gravitating system collisionlessly. 
\item To ensure that the statistical nature of the DARKexp is fulfilled
each particle must undergo numerous interactions (with the potential) during relaxation. 
\item It should allow unimpeded energy exchange between the particles and the potential.
\item It should not allow the exchange of angular momentum, or $J^2$, because DARKexp does 
not treat the redistribution of $J^2$, and it is possible, at least in principle, that in 
the process of redistributing $J^2$, the energy distribution will also be affected. 
\item The final halos should be close to isotropic. 
\end{enumerate}

The first two criteria are satisfied by the collisionless dark matter N-body simulations, 
and the fifth one can be fulfilled by selecting nearly isotropic systems from the entire 
set of virialized halos. In fact, most simulated halos are close to isotropic, except for 
the outermost radii. The third criterion is probably satisfied, but the fourth is not 
satisfied by standard N-body simulations, because they have tangential forces and hence 
allow transfer of angular momentum. For example, formation of bar-like structures or 
central triaxiality is common in simulations \citep{ma85,bel08}, and these tend to transfer 
angular momentum from the inner to the outer halo. A numerical scheme that does satisfy the 
fourth criterion is the extended secondary infall model, or ESIM, described in detail in 
\cite{wbd04} and \cite{austin05} and summarized below.

\subsection{Extended Secondary Infall Model}\label{ESIMsum}

ESIM is a geometrically spherically symmetric shell code, where each shells' angular and 
radial actions are held constant throughout the collapse. The initial conditions, detailed 
below, need not be cosmological, because the relaxation principles we are investigating are 
general, and do not hinge on cosmology. The only shell property that is allowed to change 
during evolution is its energy. ESIM does not calculate forces, instead, at every time step 
it recalculates each shell's energy, i.e., solves the equation,
$E=\Phi(r)+\frac{1}{2}\Bigl[v_t^2(r)+v_r^2(r)\Bigr]$ so that it agrees with the halo's 
potential, which is the sum of the contributions of all the individual shells. Thus, energy 
redistribution in a collisionless fashion is allowed, while angular momentum redistribution 
is not, thereby fulfilling criterion (4) of DARKexp models.

A proto-halo is divided into concentric shells; the collapse proceeds inside-out with the 
innermost shell detaching from the Hubble flow first, reaching turn-around and collapsing. 
The second shell follows, etc.\ until some final epoch is reached. As each of the hundreds 
of shells reaches turn-around and collapses, the potential changes and so the energies of 
all the interior shells have to be readjusted to satisfy the energy equation. This repeated 
shell energy readjustment is what satisfies criterion (2) of Section~\ref{chosing}.  We note 
that unlike many analytical secondary infall models, the ESIM collapse factor, i.e., the 
ratio between the turn-around radius and the final equilibrium radius for a given shell is 
not pre-set; the shell finds its apocenter and pericenter that satisfy the energy equation.


There are two types of input for ESIM: the density profile for the initial proto-halo at an
early epoch, and the {\it rms} of the random velocity dispersions. In the original \cite{rg87} 
prescription the density profile shape function is taken from \cite{bbks}, and depends on the
matter power spectrum, $P(k)$.  The {\it rms} of the random velocities are calculated based 
on $P(k)$. 

In this paper the two types of inputs are not derived from a cosmological $P(k)$. 
For a given shell the magnitude of the random velocity is picked randomly from a Maxwellian
distribution having the specified {\it rms}, and the splitting between radial and tangential
directions is done randomly. These random velocity components are imparted to the shell at
turn-around, $r_{to}$. The tangential component gives the shell's angular momentum per unit 
mass, $J=v_{tan}r_{to}$, and the radial component is added to the radial velocity developed 
through collapse. 

To test DARKexp, we need nearly isotropic systems. ESIM halo evolution drives halos toward
isotropy, but full isotropy usually cannot be achieved, because of the constraint 
that all shells keep their $J$'s fixed. To satisfy criterion (4) of Section~\ref{chosing} we 
generate a large number of halos by varying the two initial conditions, the density and 
{\it rms} velocity dispersion profiles of proto-halos, evolving them, and then selecting the 
final halos that are close to isotropic. We did not try to generate systems that would 
resemble N-body halos (or have $\phi_0$ near 4), instead we wanted our halos to span the 
broadest range of DARKexp profile behavior.

\begin{figure}[t]
\plotone{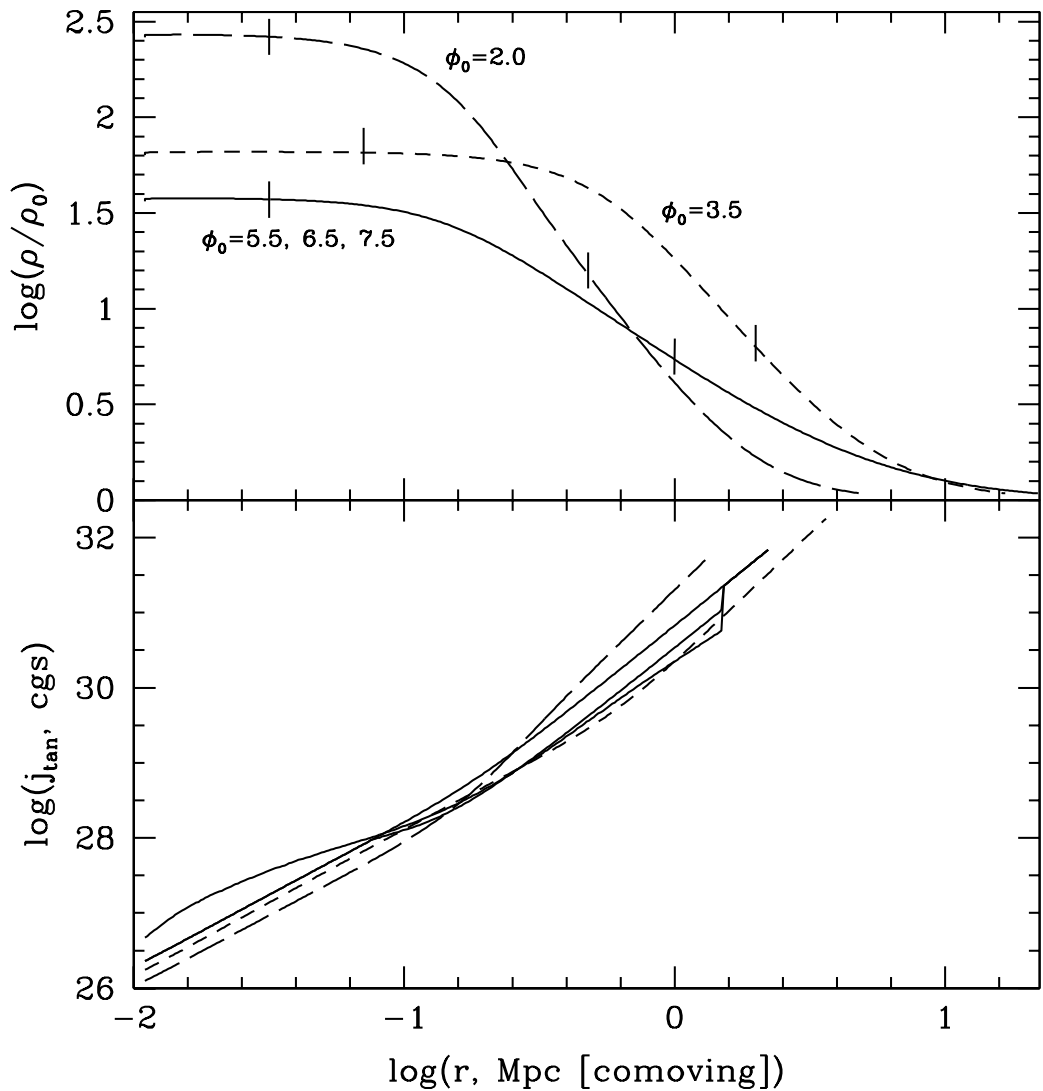}
\vskip-1.5in
\caption{Initial density profiles and angular momentum distributions for the five ESIM halos.
In the upper panel the profile are labeled by the final $\phi_0$. The initial density profile 
shown with the solid line was evolved using three different initial angular momenta profiles
(all shown with the solid line in the bottom panel). The density profiles represented by the 
long and short dashed lines in the upper panel correspond to $J(r)$ distributions of the same 
line type in the lower panel. The short vertical line segments in the upper panel indicate 
roughly the radial range shown in Figures~\ref{lnrgamma2}-\ref{esimDARK2601}.}
\label{props}
\end{figure}

We use five ESIM halos; their initial conditions are shown in Figure~\ref{props}. The top
panel has the density profiles, while the bottom has the angular momentum distribution, $J(r)$.
The density profiles represented by the long and short dashed lines in the upper panel 
correspond to $J(r)$ distributions of the same line type. The solid line density profile
was evolved using three different $J(r)$'s represented by solid lines in the bottom panel.  
Three of the initial $J(r)$ profiles (long dashed, short dashed, and one of the solid lines)
follow the original \cite{rg87} prescription, while the remaining two solid line $J(r)$'s 
are modified to generate intermediate values of $\phi_0$ in the final halos. 

Before we proceed, we point out that most physical collapse and relaxation schemes designed 
to test DARKexp predictions will be limited to some degree. The limitations of N-body simulations
were already pointed out. In ESIM, it is impossible to
strictly comply with criteria (3) and (4) of Section~\ref{chosing} simultaneously. 
Since the halos are not allowed to redistribute their angular momentum among shells, 
this indirectly imposes restrictions on how well the energy can be redistributed. For example, 
a shell which was endowed with a large $J$ as its initial condition cannot come close to the 
halo center, and thus cannot have energy arbitrarily close to $\phi_0$. Consequently, energy 
cannot be redistributed completely freely, and condition (3) may not be fully satisfied in 
all halos.

\subsection{Energy distributions}\label{energy}

\begin{figure}[t]
\plotone{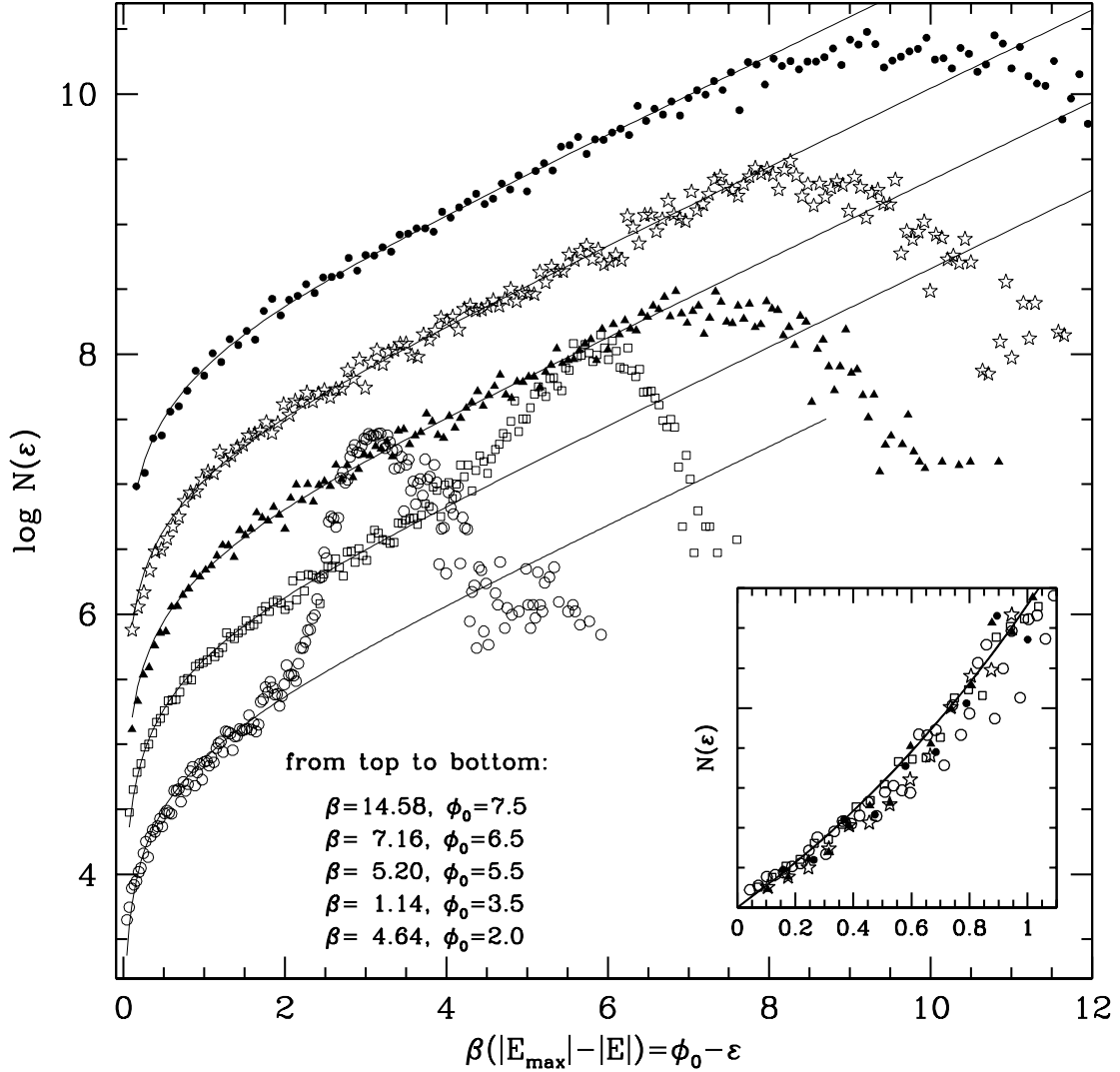}
\vskip-1.5in
\caption{Differential energy distribution of five ESIM halos. The thin lines are the best fit
DARKexp models, fit up to the energy where ESIM halos begin to clearly deviate form exponential.
The inverse temperature, $\beta$, and the dimensionless potential depth, $\phi_0$ for each halo 
are shown in the plot. The inset is a linear plot of the most bound portions of the five halos, 
up to $\phi_0-\varepsilon\sim 1$. The vertical scale has been rescaled so that all the DARKexp 
fits coincide, and are shown as the single line in the inset. Note that the most bound portions 
of ESIM halos have linear $N(\varepsilon)$.} 
\label{esimNE}
\end{figure}

Figure~\ref{esimNE} shows the energy distributions of five evolved ESIM halos described above. 
Each halo was fit with DARKexp $N(\varepsilon)$, with two free parameters, the normalization and 
temperature, $T=1/\beta$, and the fits are shown as curves.\footnote{We use capital symbols, such as
$E$ and $\Phi$, and also $T^{-1}=\beta$, to denote dimensional quantities, and lower case symbols, 
such as $\varepsilon$ and $\phi$, for dimensionless quantities.} The normalization parameter is 
related to the total mass of the halo, and is irrelevant here. In the Figure the normalization 
was set arbitrarily, to space out the curves evenly. The only relevant parameter is $\beta$.
The fitting is done over the energy range $E_{max}$ to $E_t$, where $E_{max}$ is the energy
of the most bound shell, and $E_t$ is the energy where ESIM halos begin to deviate from
a straight line in the $\log[N(E)]$ vs. $E$ plot. This choice is somewhat, but not too arbitrary;
Figure~\ref{esimNE} shows that for any given halo the deviation starts relatively abruptly.
For plotting purposes, the energy scale of each halo was multiplied by $\beta$, so the horizontal 
axis is in dimensionless units of $\beta(|E_{max}|-|E|)=\phi_0-\varepsilon$, and $\beta$ is 
different for each halo.

It is apparent that ESIM energy distributions are very well fit by DARKexp. The deviations are 
seen at large $\phi_0-\varepsilon$, i.e., for loosely bound shells which inhabit the outer, 
not yet equilibrated portions of halos. For most ESIM halos the DARKexp fit is valid for a wide 
enough range of energies and $N(\varepsilon)$, so that the best-fitting function is, unmistakably, 
$N(\varepsilon)\propto[\exp(\phi_0-\varepsilon)-1]$. 

\begin{figure}[t]
\plotone{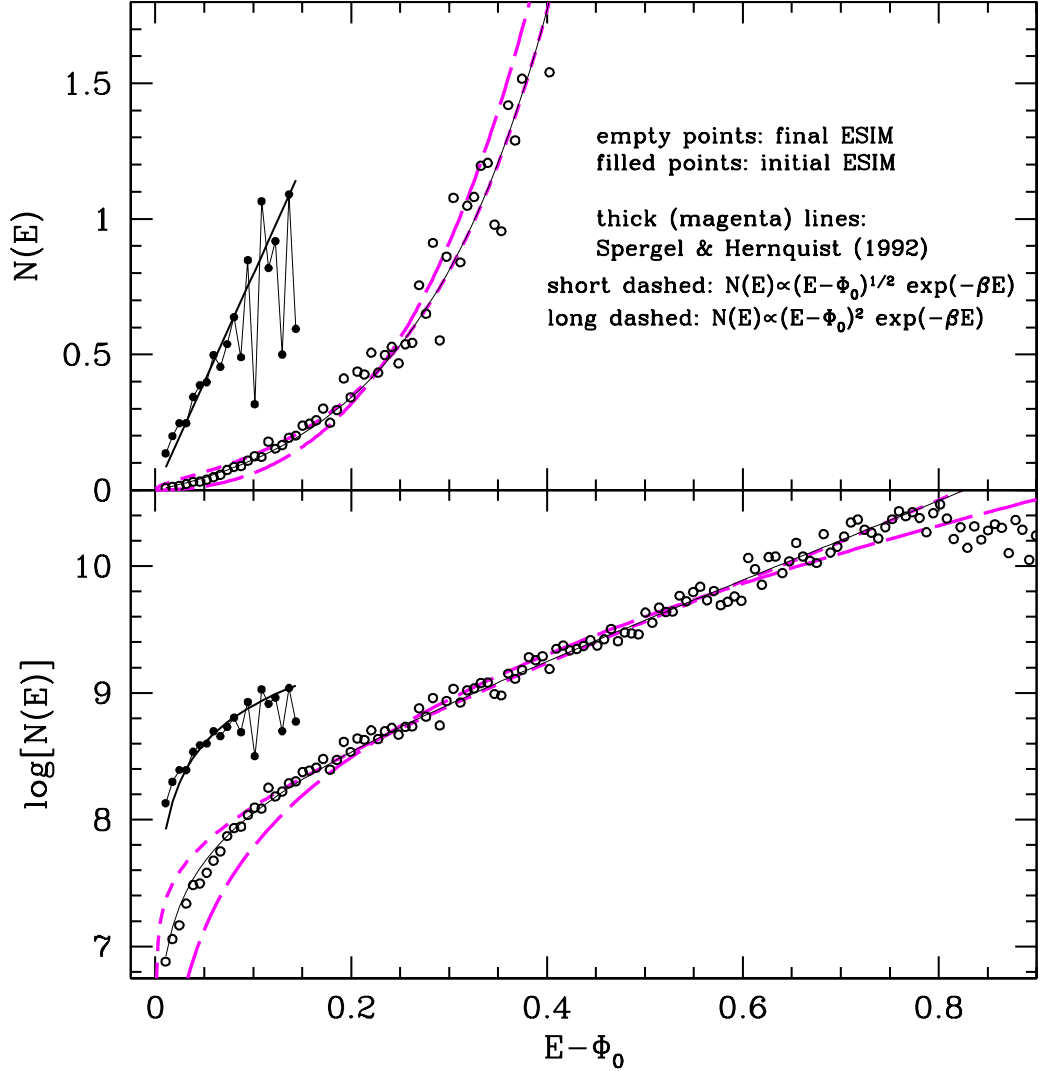}
\vskip-1.5in
\caption{The $\phi_0=6.5$ ESIM halo of Figure~\ref{esimNE}. The top panel has a linear vertical 
scale, while the bottom panel, log. The filled points are the initial, pre-collapse energy 
distribution, and the empty points are the final. (Vertical normalization is arbitrary.)
Note that the initial energy distribution is narrow, has a shallower $\Phi_0$ than the final 
distribution, and is roughly linear in energy. The two thick magenta lines represent the two 
extreme cases of the \cite{sh92} model, as labeled in the plot.
}
\label{smNEif9811}
\end{figure}

To illustrate this point we compare in Figure~\ref{smNEif9811} the $\phi_0=6.5$ ESIM halo (empty
points) from Figure~\ref{esimNE} to the prediction of the scattering model proposed by \cite{sh92}. 
Their $N(E)$ goes approximately as $[(E-\Phi_0)^{-2}+C(E-\Phi_0)^{-1/2}]^{-1}\exp(-\beta E)$. 
The two extremes of this function are shown as the short dashed (magenta) line in the limit where 
$N(E)=(E-\Phi_0)^{1/2}\exp(-\beta E)$, and long dashed (magenta) line in the limit where 
$N(E)=(E-\Phi_0)^{2}\exp(-\beta E)$. Both fail to match ESIM halos.\footnote{It is not clear 
how parameter $C$ was obtained in their analysis. Inspection of their Figure 2 shows that the 
$N(E)=(E-\Phi_0)^{1/2}\exp(-\beta E)$, case is more appropriate. If parameter $C$ can be chosen 
arbitrarily, then good fits to ESIM halos are possible, as might be expected because the model 
has one more degree of freedom.} 

We also compare DARKexp to another physically motivated model of halo formation. Non-extensive 
statistical mechanics \citep{t88} predict collisionless systems to be polytropes \citep{pp93}, 
which have distribution functions that are power laws in energy, with a cutoff at the most bound 
energy (\cite{BT87}, eq. 4-105).  The distribution function of the central regions of DARKexp 
is also a truncated power law (Paper I, eq. 21), but the exponent is $n=-1$, while polytropes 
have $n>1/2$. In the outer regions both polytropes and DARKexp density profiles are 
approximately power laws in radius, but the transition between the inner and outer slope 
is much faster in polytropes than in the DARKexp systems. Overall, DARKexp is quite different 
from polytropes.

Note that the initial, pre-collapse shape of the ESIM energy distribution is very different 
from the final, so the goodness of the DARKexp fit is not the result of (un)luckily chosen 
initial conditions. The filled points in Figure~\ref{smNEif9811} show the initial energy 
distribution of the $\phi_0=6.5$ ~ESIM halo. It is much narrower than the final distribution 
and is linear in energy; also the potential depth of the pre-collapse halo, $\Phi_0$, is 
considerably shallower than that of the final. The initial conditions of the other four
halos are similar. 

Because ESIM is a shell code with the proto-halo shell thickness increasing outward 
logarithmically, the mass resolution is considerably better in the inner regions than in 
the outer. The portions of the energy distribution histogram with energies near most bound 
typically have a few hundred (relatively low mass) shells in each bin, so the region in 
$N(\varepsilon)$ which matters most for comparison with DARKexp has the highest resolution. 
This is not so for N-body codes, where each particle has equal mass. The inset in 
Figure~\ref{esimNE} shows that near the most bound energies the distribution is linear,
as is expected from DARKexp, 
$N(\varepsilon)=\exp(\phi_0-\varepsilon)-1\approx 1+(\phi_0-\varepsilon)+(\phi_0-\varepsilon)^2/2+... -1\approx (\phi-\varepsilon)$.

\subsection{Density, velocity dispersion, and $\rho/\sigma^3$ profiles}

In addition to the energy distribution, one can also do the comparison with the density and
velocity distributions. Because DARKexp $N(\varepsilon)$ is 
dimensionless (for example, the virial radius, $r_{200}$ has no meaning in DARKexp), and ESIM 
quantities are dimensional ($r_{200}$ is clearly defined), we first have to put these on the 
same footing. To that end, we estimate the dimensionless potential depth $\phi_0$ 
for each ESIM halo using $\phi_0=\beta(|E_{max}|-|E_{min}|)$, where $E_{min}$ is the energy of 
the least bound shell. What energy represents the least bound shell is not clear; potential 
differences are readily known, but not absolute potential values. We estimate $|E_{min}|$ as 
$|E_{200}+E_{t}|/2$, where $E_{200}$ is the typical energy of particles at the radius where 
the average interior density is 200 times the critical, and $E_t$ is defined in 
Section~\ref{energy}. Note that $|E_{200}|$ is always less than $|E_t|$, implying that ESIM 
halos are not equilibrated up to the virial radius. 

\begin{figure}[t]
\plotone{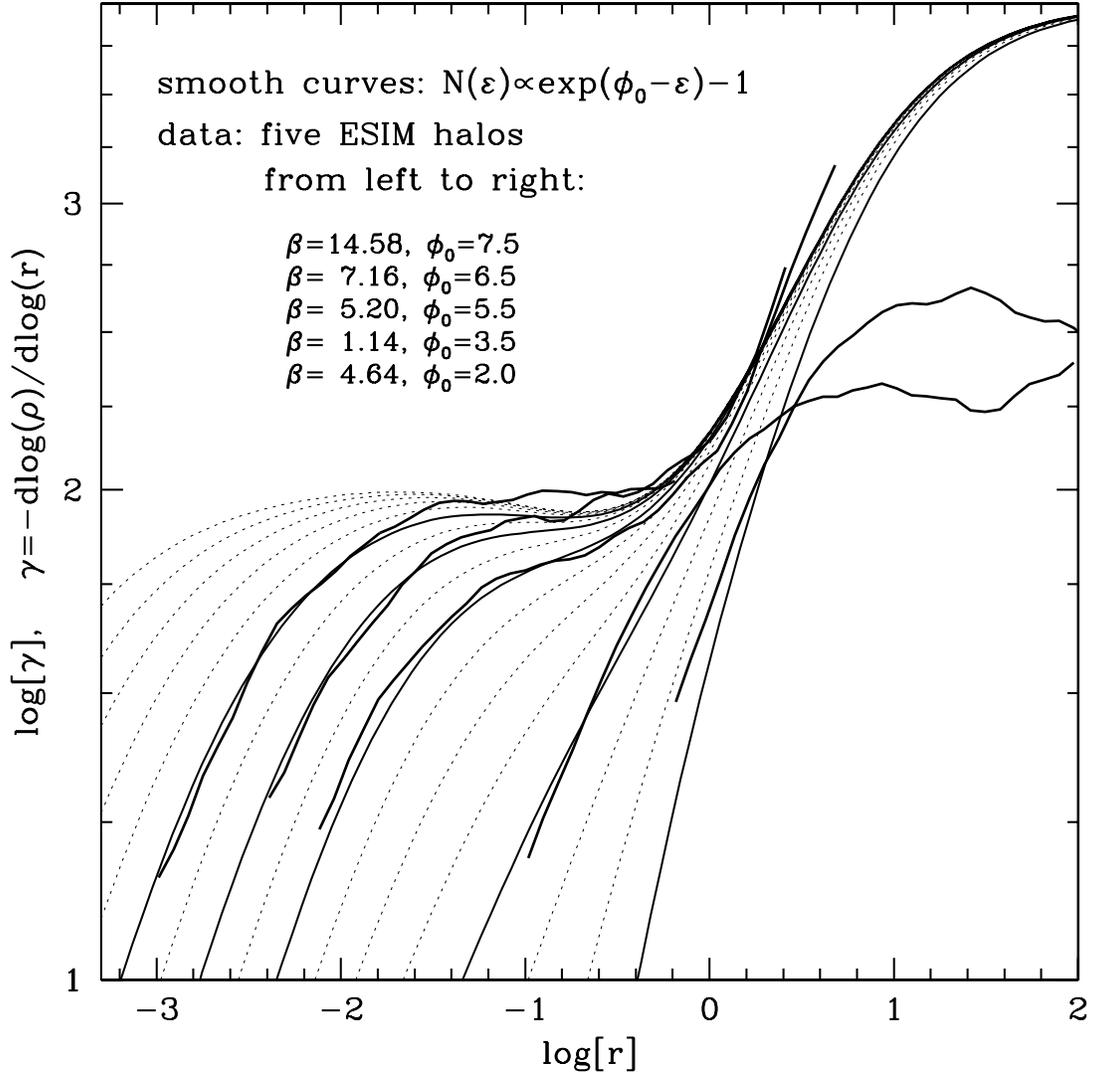}
\vskip-1.5in
\caption{Similar to Figure~\ref{slopesALL}, but with five ESIM halos superimposed. The corresponding
DARKexp models are highlighted as thick lines; the rest of DARKexp models (from $\phi_0=2$ to 10,
in steps of $0.5$) are shown as thin dotted lines. Compared to Figure~\ref{slopesALL}, in this 
Figure we spaced out the curves horizontally, to avoid overcrowding.} 
\label{lnrgamma2}
\end{figure}

Given this estimate of $\phi_0$, we can check how well the ESIM density profiles reproduce the 
corresponding DARKexp $\rho(r)$. The two should match reasonably well, but perfect agreement 
is not expected because (i) some ESIM halos do not follow DARKexp $N(\varepsilon)$ for a wide 
energy range (see Section~\ref{ESIMsum}), (ii) estimating $E_{min}$ is not rigorous, and (iii) 
none of the ESIM halos are perfectly isotropic. 
All five halos are shown in Figure~\ref{lnrgamma2}, against the background of DARKexp models
with  $\phi_0$ values from 2 to 10, in steps of 0.5; the ones fitting the five ESIM halos are
shown as thicker lines.  As for $N(\varepsilon)$, the DARKexp models provide very good fits to 
most of the ESIM halo density profiles. ESIM halos with the shallower potentials do not match 
DARKexp predictions as well as the ones with deeper potentials. This could be because the 
shallower halos do not undergo as much change in the potential during the evolution, and hence
do not relax fully. 

\begin{figure}[t]
\plotone{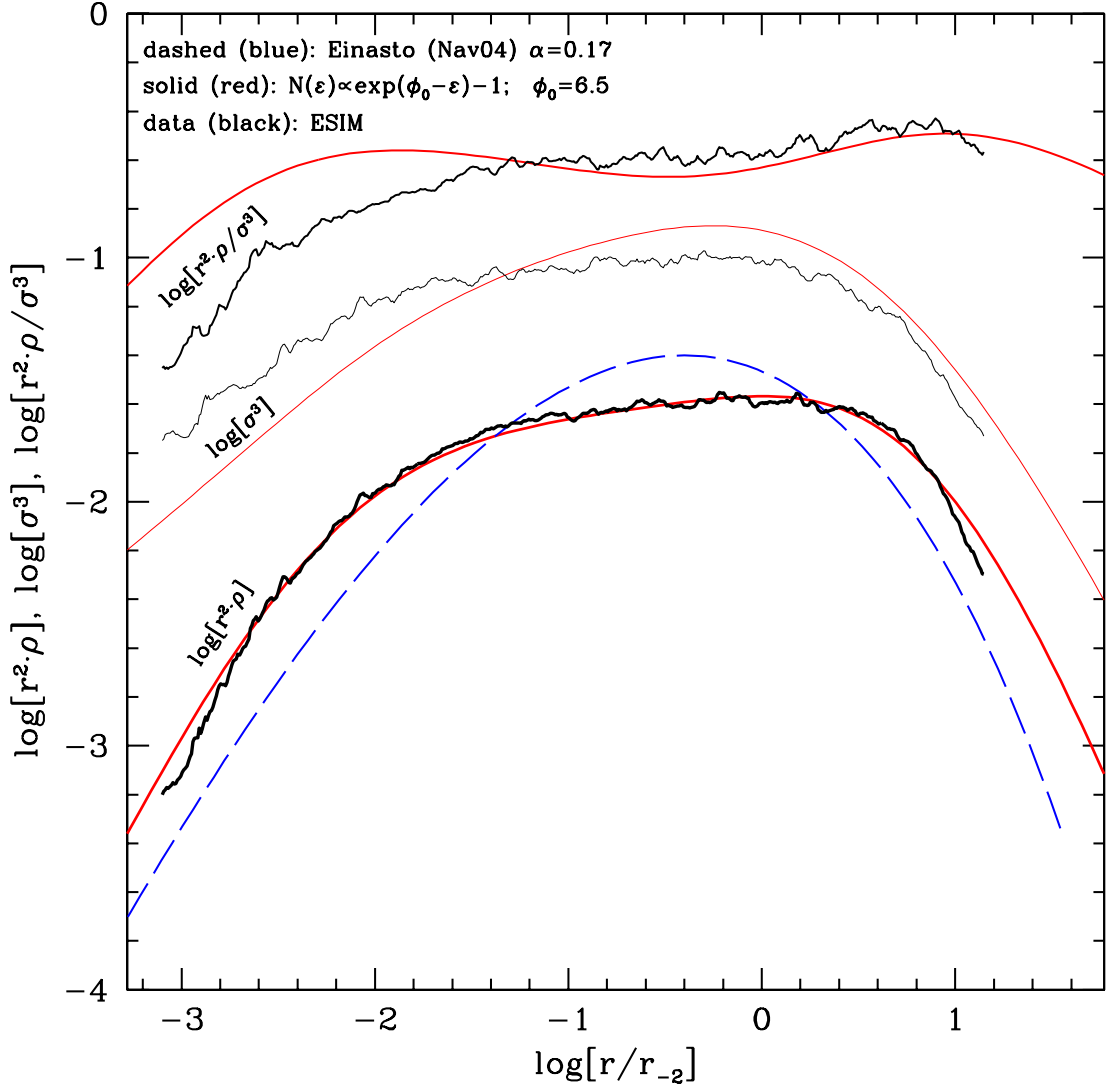}
\vskip-1.5in
\caption{Density, velocity dispersion, and pseudo phase-space density of the ESIM (black) and the 
corresponding DARKexp halo (red), with $\phi_0=6.5$ plotted in Figure~\ref{esimNE}. The lowest two 
curves (thick lines) are the density profiles, multiplied by $r^2$. The DARKexp is an excellent 
fit to ESIM over 4 decades in radius. The dashed line is the Einasto (or, \cite{nav04}), shown for 
comparison. The middle two curves (thin lines) are: 
$\sigma^3$ for the isotropic DARKexp models, and $\sigma^3=\sigma_{rad}\sigma_{tan}^2$ for ESIM, 
which are not exactly isotropic. The difference between ESIM and DARKexp velocity dispersion profiles
are due to ESIM halo anisotropy. This difference is also reflected in the top set of two curves,
$r^2\cdot(\rho/\sigma^3)$. The vertical normalization of curves is arbitrary.} 
\label{esimDARK9811}
\end{figure}

\begin{figure}[t]
\plotone{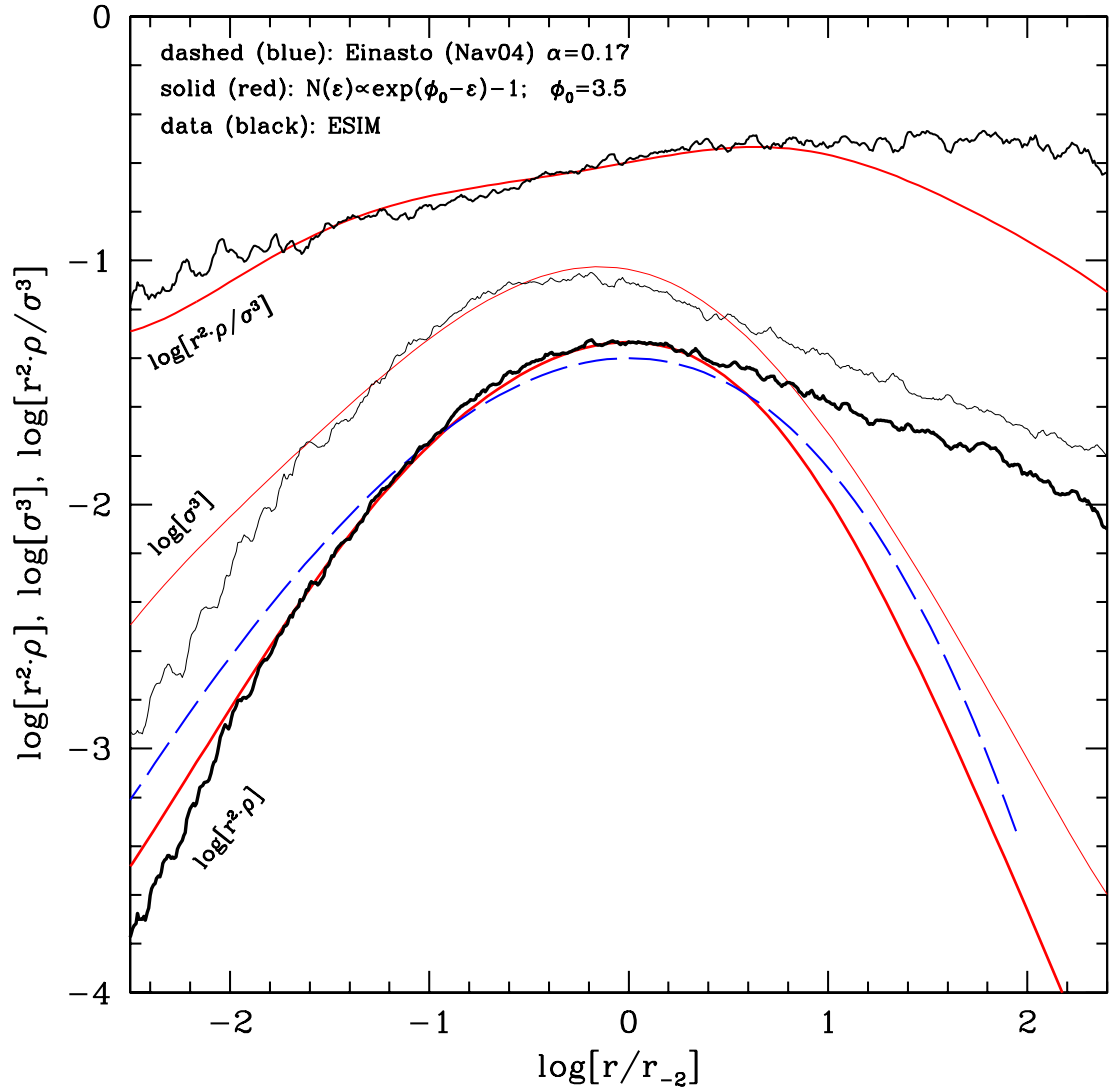}
\vskip-1.5in
\caption{Same as Figure~\ref{esimDARK9811}, but for the ESIM and the corresponding DARKexp halo 
with $\phi_0=3.5$. Note that here the ESIM density and the velocity dispersion profiles deviate 
from DARKexp, possibly because the former is not fully relaxed owing its shallow potential. 
ESIM $\rho/\sigma^3$ remains a power-law.} 
\label{esimDARK2601}
\end{figure}

The second ($\phi_0=6.5$) and the fourth ($\phi_0=3.5$) of the five profiles of Figure~\ref{esimNE} 
are shown in Figures~\ref{esimDARK9811} and \ref{esimDARK2601}, where we plot $\log(\rho\cdot r^2)$ of 
the ESIM halos (black lines with 'noise'), and the best-fitting DARKexp curve (red smooth lines). 
The radii are in units of $r_{-2}$, but the vertical normalization is arbitrary, so only the shapes 
of the curves are to be compared. 
The halo with the deeper potential (Figure~\ref{esimDARK9811}) closely follows the DARKexp 
prediction for nearly four decades in radius, whereas the shallower one, does so for only two decades, 
and deviations become pronounced in the outer halo. We note that these density profile shapes, 
especially for the deeper potentials, like in Figure~\ref{esimDARK9811}, where the density slope
stays roughly constant for a finite radial range were already evident in Figure 1 of \cite{wbd04}.

We use the isotropic Jeans equation to obtain the velocity dispersion profile for each halo, shown 
as thin lines in Figures~\ref{esimDARK9811} and \ref{esimDARK2601}. Note that the DARKexp~ $\sigma^3$
shape tracks the density profiles shape, ensuring that $\rho/\sigma^3$ (the top smooth line)
stays close to a power law for $\sim 3$ decades in radius. The reason for the pseudo phase-space
density, $\rho/\sigma^3$ being a power law in collisionless simulations is still not known, but it
is reassuring that DARKexp models predict this feature. 

In Figure~\ref{esimDARK9811}, while the ESIM density profile is very well fit by DARKexp, the 
shape of the velocity dispersion profile, $\sigma^3$, which we define as 
$\sigma_{rad}\sigma_\theta\sigma_\phi=\sigma_{rad}\sigma_{tan}^2=\sigma_{rad}^3[1-\beta(r)]$,
deviates from DARKexp predictions because of ESIM's non-zero anisotropy, $\beta(r)$. The top set 
of curves shows $\rho/\sigma^3$, which also deviate somewhat from DARKexp predictions, again, 
because of anisotropy.

\section{Conclusions}

Because the sole dynamical action of the ESIM is to repeatedly
readjust shells' energies in response to the changing potential, it is well suited for 
testing the DARKexp predictions described in Paper I \citep{hw10}.
This constant reshuffling of energies drives halos to the most probable, DARKexp state.
The more realistic numerical schemes, like the N-body simulations are also more complex
in the sense that they do more than just redistribute particle energies, they also affect 
the redistribution of angular momentum. Whether the inclusion of this additional degree 
of freedom will preserve the DARKexp $N(\varepsilon)$ form is not yet clear. We speculate
that the changes, at least in the inner portions of halos will be minor because N-body 
halos are fully relaxed and isotropic in those regions.

The argument presented in Paper I does not rely on any specific process, like infalling 
substructure or dynamical friction, to achieve the final state; it is a maximum entropy argument. 
If the same principle applies to the redistribution of angular momentum in collisionless 
simulations of dark matter halos, then the final $N(E,J^2)$ distribution can also be arrived at 
similarly, without knowing the details of the radial orbit, or other tangential instabilities
that operate in simulations. Encouraged by the success of our isotropic investigations we will 
pursue the statistical mechanical approach, and apply it to anisotropic systems.

\acknowledgments
The Dark Cosmology Centre is funded by the Danish National Research Foundation. 
L.L.R.W. is very grateful for the hospitality of the Dark Cosmology Centre at the University
of Copenhagen, and the Institute for Theoretical Physics at the University of Z\"urich, 
where she spent the Fall of 2009 and the Spring of 2010, respectively.

\end{document}